# Interaction of 160 GeV- Muon with Emulsion Nuclei

S. M. Othman(1), M.T. Ghoneim(2), M.T. Hussein(2),

H. El-Smman(1) and A. Hussein(1)

(1) Physics Department, Faculty of Science, Menufia University, Egypt

(2) Physics Department, Faculty of Science, Cairo University, 12613, Giza, Egypt

## ABSTRACT

In this work we present some results of the interaction of high energy muons with emulsion nuclei. The interaction results in emission of a number of fragments as a consequence of electromagnetic dissociation of the excited target nuclei. This excitation is attributed to absorption of photons by the target nuclei due to the intense electric field of the very fast incident muon particles. The interactions take place at impact parameters that allows ultra-peripheral collisions to take place, leading to giant resonances and hence multi-fragmentation of emulsion targets. Charge identification, range, energy spectra, angular distribution and topological cross-section of the produced fragments are measured and evaluated.

## INTRODUCTION

Ultra-peripheral nuclear collisions beyond the area of hadronic interactions have been brought to the concern of several physicists in the recent decades [1-5]. The study of electromagnetic interactions at relativistic energy is, in fact, of multifold importance; they yielded valuable information about the size and structure of the nuclei as well as the properties of giant resonance viewed as collective excitations through several experiments studying the interactions of electrons and photons with nuclei [6-8]. Deep inelastic scattering off protons and nuclei has led to clear understanding of their partonic structure [9]. Electromagnetic dissociation of nuclei at intermediate energies allowed studying reactions that were the inverse of the nucleosynthesis and double giant resonances [10-13]. Electromagnetic excitation is one of few ways to investigate the fissility of unstable nuclei [14]. Thus, electromagnetic interactions have been good ways to probe many physics aspects.

Muon, being a lepton, would interact with nuclei by either weak or electromagnetic mechanism. Weizsacker – Williams's introduced a simple classical approach to describe the reaction mechanism, in which, the high energy muon particle is looked upon as a lump of intense electric field moving at a high speed [15]. The range of the impact parameter in this case is much larger than the range of the nuclear force so that it is called "ultra-peripheral" interaction. The impact of the Lorentz contracted Coulomb field of this fast moving particle on one of the emulsion target nuclei leads to absorption of one or several equivalent photons by this nucleus. The concept of equivalent photons was proposed by Fermi who considered the field of a moving charge to be viewed as a photon flux and used this concept to solve the problems of interactions between moving charges and nuclei [16-17].

It is clear that in such a type of interaction, the interacting nuclei do not penetrate each other, i.e., the target nucleus is excited by the absorption of virtual photon(s) in the electromagnetic field of the incident particles then decays by particle emission. Hoffman et al. introduced a theoretical study based on the Coulomb dissociation over a wide range of projectile energies [18]. Experimental high energy physicists have been working on the multi-fragmentation process of nuclei to provide broader understanding of nuclear structure [19-20]. In this work, we present some general results of the interaction under test. However, in forthcoming series of papers, we will present extended studies, in view of theoretical models, that come out of the data harvested from this experiment.

.

## 1-EXPERIMENTAL WORK

Nuclear emulsion pellicles of a type equivalent to BR-2 one, having dimensions of 9 cm × 12 cm and thicknesses of 60 μm were irradiated with 160 GeV -beam of muons at CERN [21]. The emulsion samples were positioned both along and across the beam. In case of the transverse orientation, a nine – hour irradiation was enough for the analysis under test. The emulsion pellicles were processed at the "Nuclotron Lab" of the Joint Institute for Nuclear Research (JINR, Dubna- Russia).

The constituents of photographic nuclear emulsions are mainly composed of H, C, N, O, Ag and Br- nuclei with relative concentrations that depend upon the type and the purpose of the experiment. The chemical composition of the emulsion in this experiment is shown in table(1).

Table (1): chemical composition of NIKFI-BR2 emulsion.

| Element | $^{1}$H | $^{12}$C | $^{14}$N | $^{16}$O | $^{80}$Br | $^{108}$Ag |
|---|---|---|---|---|---|---|
| Atom/cm$^3$ x10$^{22}$ | 3.150 | 1.410 | 0.395 | 0.956 | 1.028 | 1.028 |

Exact target identification in an emulsion experiment is not an easy task as the medium is composed of various elements as mentioned above. However, the major constituent elements can be divided into two broad target groups such as C, N, O (light targets) and Ag, Br (heavy targets). The data were collected by the area scanning technique of the nuclear emulsion plates of our study. Track counting and track parameter determination were performed using an optical microscope of model "Olympus BH-2" using 100X oil emersion objectives and 15X eyepieces. The field area is calibrated using standard slits and a fine micrometer attached to eye piece lens of 0.1 μm per division.

A total of 820 interaction events of the muon projectile with emulsion target nuclei were collected by research group workers of the high energy Lab of Cairo and JINR within the scientific collaboration between the two groups. The fragments emitted from the target were observed as ionizing particles around the vertex of fragmentation. Thickness, coordinates and range, were measured carefully and recorded for each of the emitted fragments coming out of each event.

## 2 Experimental results

### 2-1 Charge Identification:

The 820 events resulted from interactions of the 160 GeV - muons with emulsion nuclei have produced 2666 fragments that were recorded and considered in our analysis. For each fragment, widths of three sections were measured at fairly separated positions along the track and then an average value was obtained. The distribution of the track widths of different target fragments together with their best fitting curves is presented in figure (1).

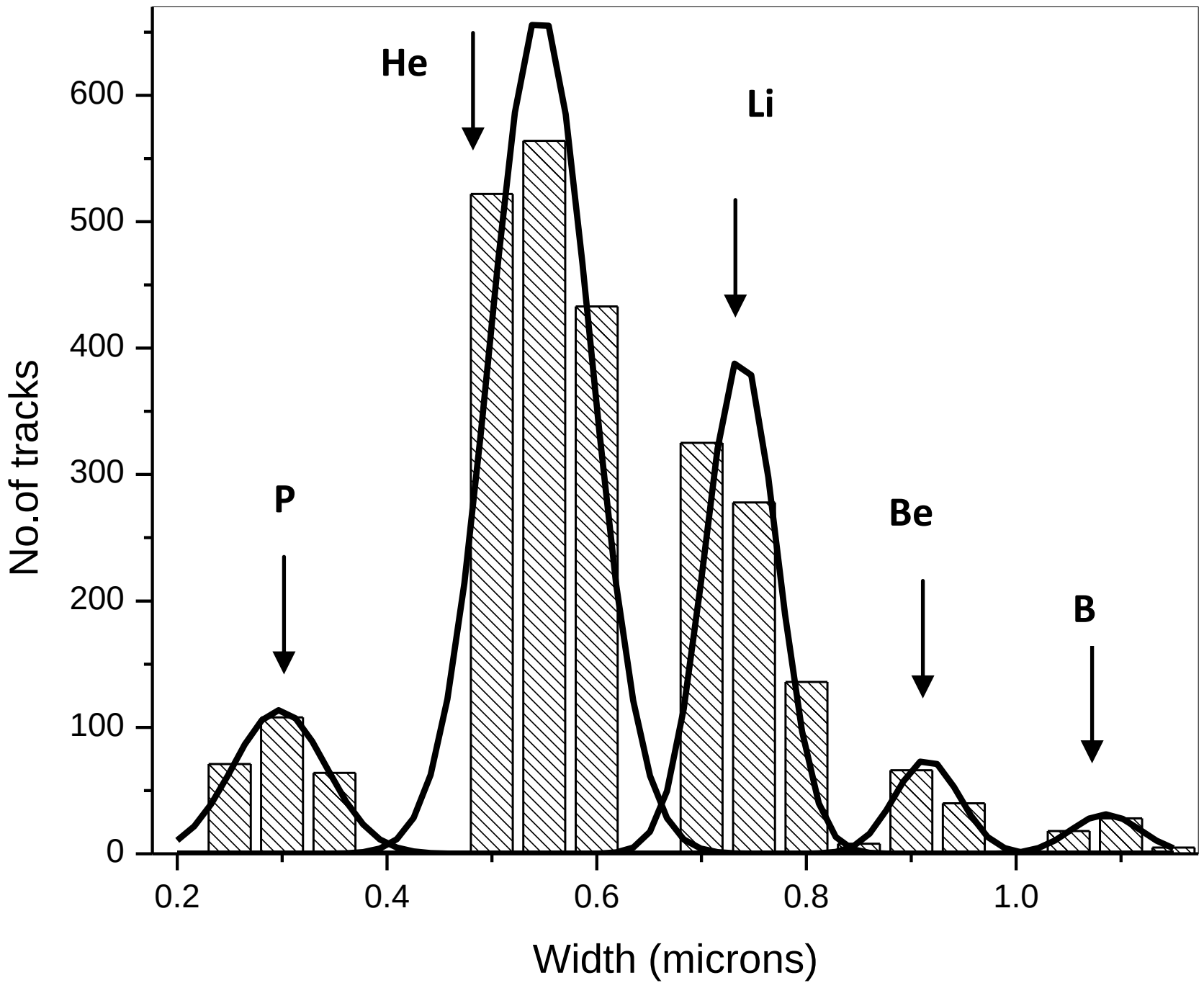

**Fig.(1):** distribution of the track widths of different target fragments. Histograms are experimental data and curves are Gaussian fitting curves.

Figure1 shows clear five distinct peaks representing different target fragments, that were fairly attributed to proton, He, Li, Be and B ionizing particles, respectively. Some carbon fragments were also distinguished, but they were not included in this graph because of their very low statistics.

The fact that emulsion targets is mainly a mixture of two groups; light nuclei (C, N and O) and heavy ones (Ag and Br), the sum of produced fragment charges, $\Sigma Z_{tf}$, in each event was used as the criterion to pick up pure interactions with the Ag, Br from all others. So, those having $\Sigma Z_{tf} > 8$ were due interaction of the muons with the heavy target group [22]. On the basis of this criterion, 253 of the whole interactions were obtained with Ag, Br target nuclei. The yield of the emitted fragments from emulsion target and from the Ag, Br ones with their percentage in each category are displayed in table (2).

Table (2): yield of the emitted fragments from emulsion (Em) targets and of Ag, Br ones.

| Fragment | Em | | Ag, Br | |
|---|---|---|---|---|
| | number of fragments | Fraction (%) | number of fragments | Fraction (%) |
| P | 243 | 9 | 49 | 5 |
| He | 1519 | 57 | 523 | 49 |
| Li | 739 | 28 | 366 | 34 |
| Be | 114 | 4 | 79 | 7 |
| B | 51 | 2 | 56 | 5 |
| Total number | 2666 | | 1073 | |

The fact that emission of light fragments comes from both light and heavy emulsion target nuclei, while emission of heavy fragments comes mostly from heavy target nuclei, gives a higher percentage of projecting out light fragments than heavy ones, as appears in table (2). This might be viewed from another point if we assume that heavier emulsion target nuclei are more exposed to be highly excited by absorbing the whole energy of the impinging virtual photons than lighter targets. This concept would lead to a development of higher temperatures in the Ag, Br nuclei and consequently, suitable conditions of "kicking out" heavy fragments. Moreover, the probability of all target nuclei to emit alpha particles seems to be remarkably large compared to emitting other fragments. This experimental fact ensures that alpha-particle is a tightly bound structure and a strong possibility that nuclei might be considered as composed of alpha clusters, as supported by several research workers [23-29].

**2-2 Angular Measurement**

The spatial configuration of each event was constructed by following up the x, y, and z coordinates of each track, to determine their dip and projection angles, $\theta_d$ and $\theta_p$ as shown in fig.2:

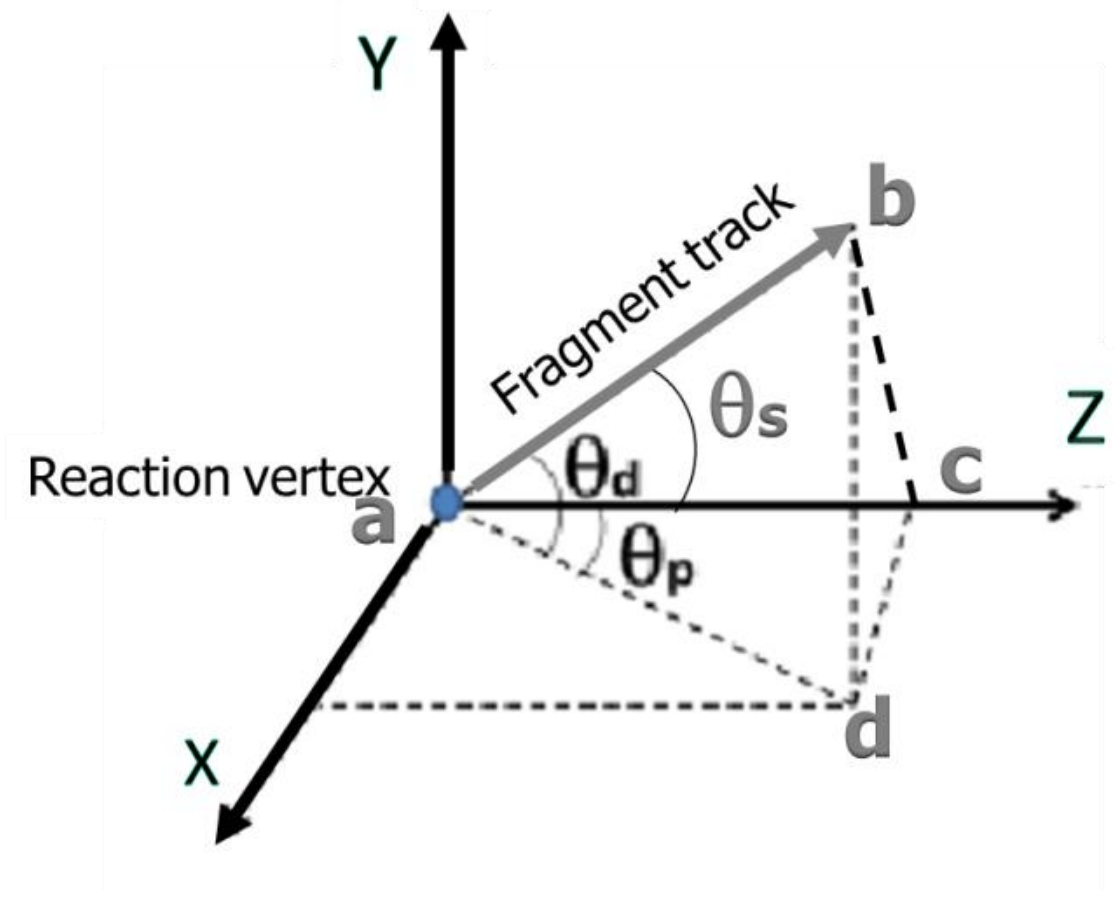

**Fig.2:** The space angle "$\theta_s$", the dip angle "$\theta_d$" and the projection "$\theta_p$" of the emitted fragments.

The Projection angle is measured as:

$$\cos\theta_p = \frac{ac}{ad} = \frac{s\,\Delta z}{\sqrt{(\Delta x)^2 + s(\Delta z)^2}} \quad (1)$$

S, is the shrinkage factor; which is the ratio between the thickness of the unprocessed and the processed emulsion and $\Delta z$ is the change in z- coordinate while travelling a distance $\Delta x$ and $\Delta y$ in the (x – y) plane.

The dip angle is simply:

$$\cos\theta_d = \frac{ad}{ab} = \frac{\sqrt{(\Delta x)^2 + s(\Delta z)^2}}{\sqrt{(\Delta x)^2 + (\Delta y)^2 + s(\Delta z)^2}} \qquad (2)$$

The space angle $\theta_S$ between the track and the z- axis (direction of projectile incidence) is thus given by:

$$\theta_s = cos^{-1}\,[cos\,\theta_p\,.\,cos\,\theta_d] \qquad (3)$$

So, by measuring the x, y and z– displacements along the linear portion of the emitted track, fairly near the interaction, we were able to get the angular distributions of the produced fragments. The angular distribution of the emitted fragments, protons through B, by emulsion nuclei and the Ag, Br group of the emulsion, are shown in Figure 3a, b, c, d and e, respectively.

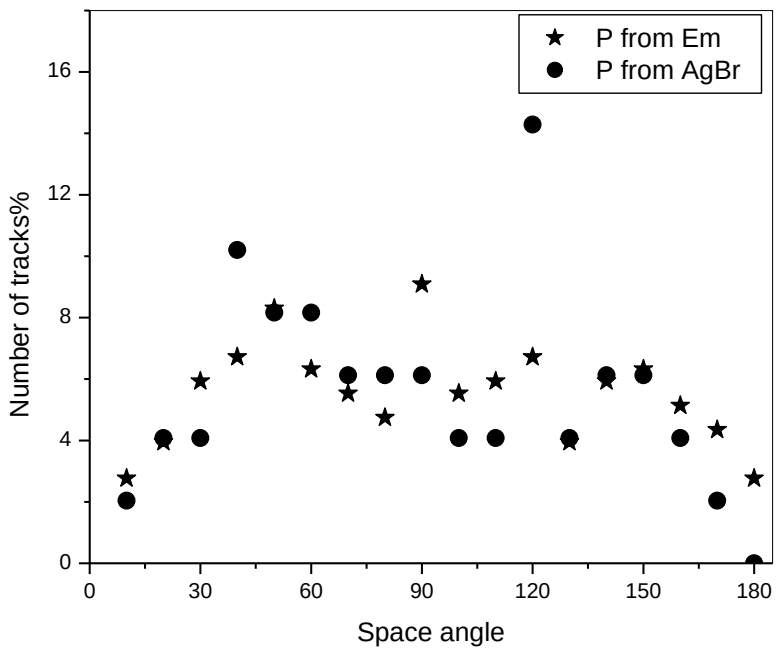

(a)

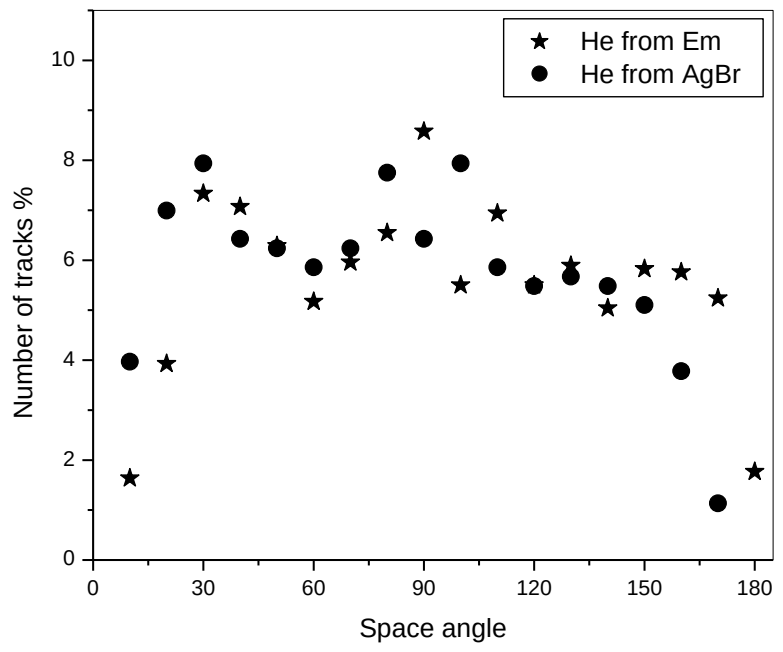

(b)

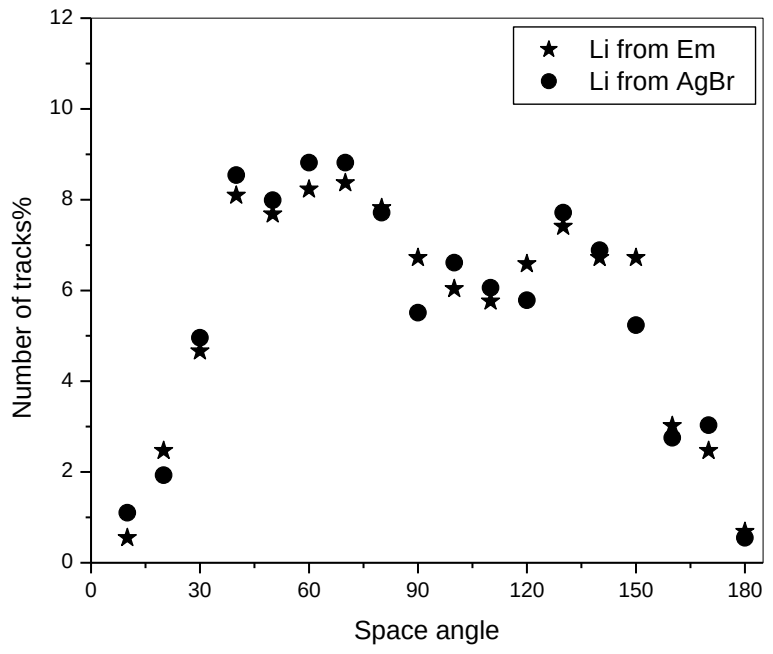

(c)

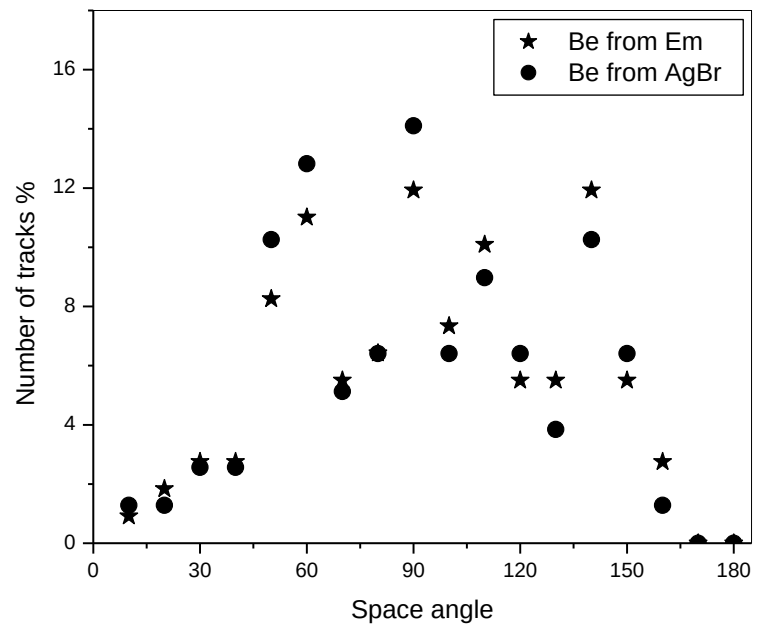

(d)

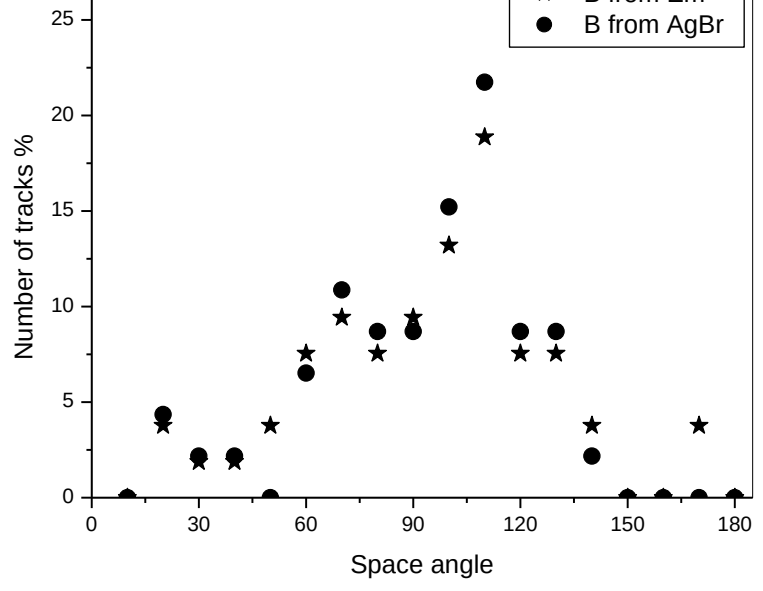

(e)

Fig.3:Angular distribution of the emitted fragments in the interactions of muon with emulsion nuclei and Ag, Br group.

In figure 3, one may see that emission of lightest fragment, the proton, is distributed over a wide range of angles. The range of the emission angles seems to get narrower and narrower as the emitted fragment gets heavier, as shown by figures 3a through 3e. This effect seems also to be the same regardless of the emitting target size, indicating a similar evaporation mechanism for the fragments for all target nuclei [30]. One may also notice that some peaks becomes one peak only in the distribution of the heaviest fragments of the Boron, figure 3e. This result refers to the fact that light fragments may be produced at different temperatures of the emitting system in earlier stages before approaching the equilibrium states, while emission of the Boron took place after the system has reached a complete evaporation state with unique boiling off temperature.

A quantitative description of the angular emissions presented in figure 3 (a - e) might be seen by the asymmetry factor, A. This factor is a measure of the anisotropy of the emitted fragments, given as:

$$A = \frac{F - B}{F + B}$$

Where, F and B are the number of fragments emitted in the forward hemisphere and backward ones. The values of this factor are presented in Table 3.

Table 3: Asymmetry factor for the emitted fragments by emulsion (Em) and Ag, Br targets.

| Fragment | A | |
|---|---|---|
| | **Emulsion** | **Ag, Br** |
| p | **0.070 $\pm$ 0.010** | **0.10$\pm$0.01** |
| He | **0.050 $\pm$ 0.008** | **0.09$\pm$0.01** |
| Li | **0.090 $\pm$ 0.010** | **0.130$\pm$0.020** |
| Be | **0.030 $\pm$ 0.001** | **0.130$\pm$0.018** |
| B | **-0.090 $\pm$ 0.009** | **-0.130$\pm$0.017** |

The asymmetry factor in table 3, does not show any preferable emission direction whether to forward or backward.

**2-3 Energy of the Fragments:**

The residual range, R, which is the path length required to bring the fragment to rest after its emission was used for energy measurement. R is given as:

$$R = (\Delta x^2 + \Delta y^2 + S^2 \Delta z^2)^{1/2} \qquad (4)$$

Then; we obtained the energy of the emitted fragments by making use of a range - energy relation that is valid for a wide energy interval [31]:

$$E = \alpha \left(\frac{m}{m_p}\right)^{1-n} Z^{2n} R^n \qquad (5)$$

In this formula; Z, $m_p$, m, are the fragment charge, the proton mass and the mass of the fragment; respectively, while α and n are constants of values 0.25 and 0.58, respectively. Our data were logged into SRIM computer program to calculate the energies of the emitted fragments. This program works over a relative wide energy 10 eV/amu up to 2 GeV/amu and uses a full quantum mechanical treatment of ion-atom (medium) collisions [32]. The energy spectra of all the fragments emitted in the disintegration of all emulsion nuclei (Em) or from heavy target only (AgBr) are given in figure 4.

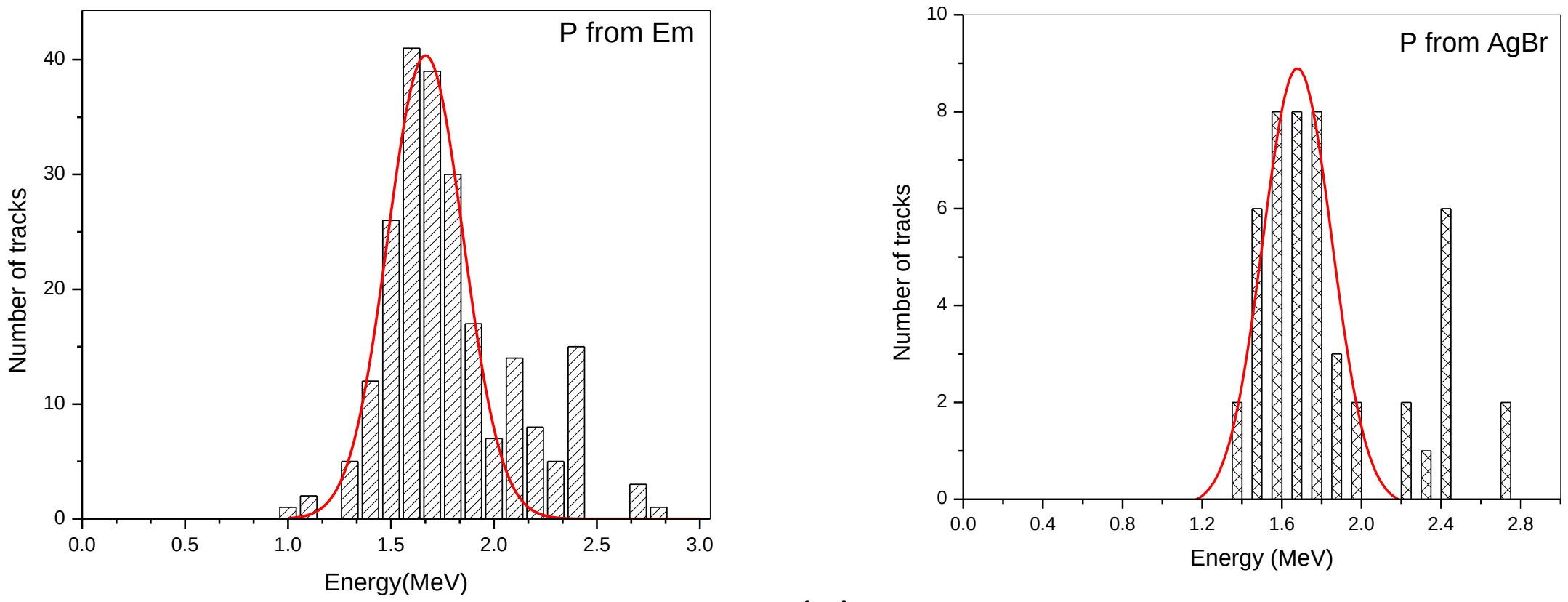

(a)

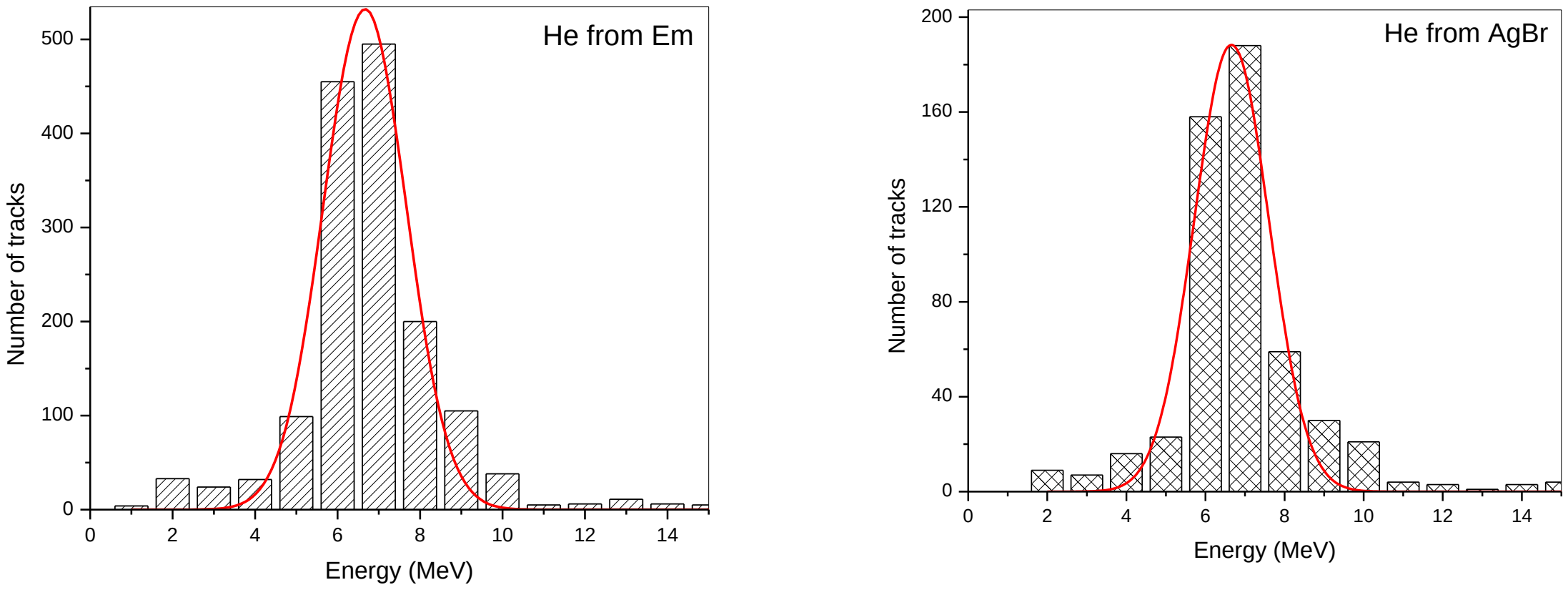

(b)

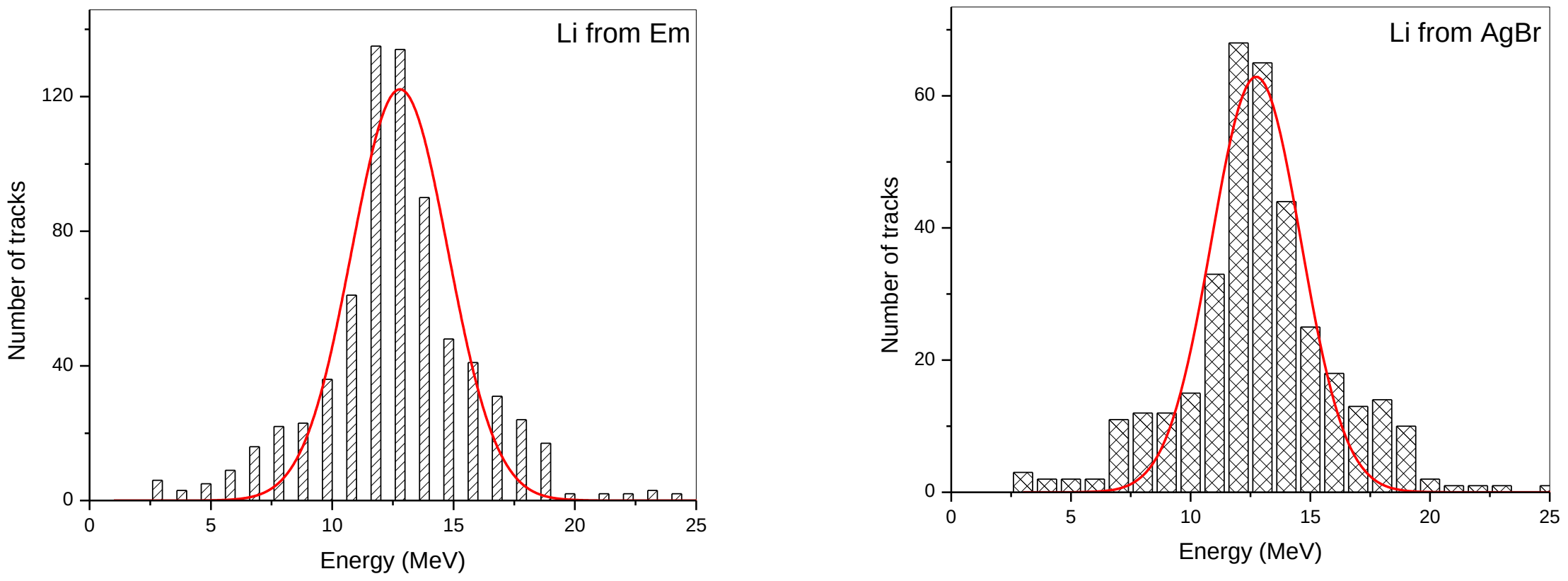

(c)

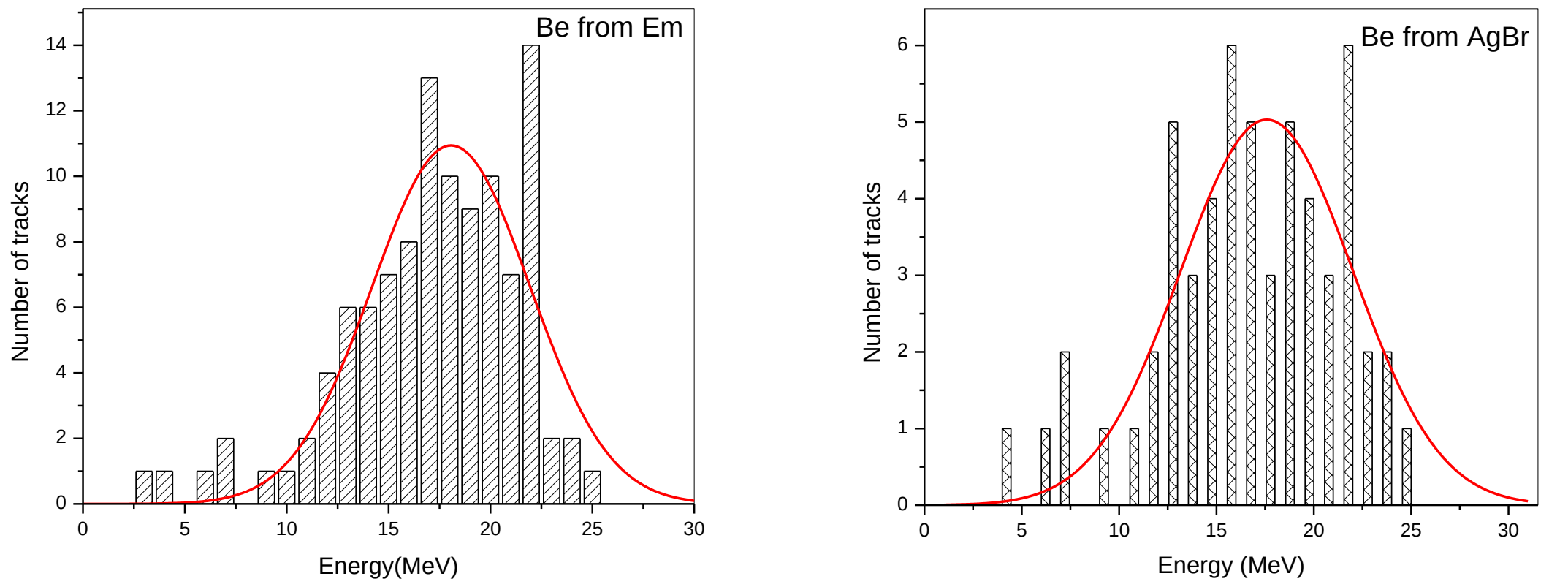

(d)

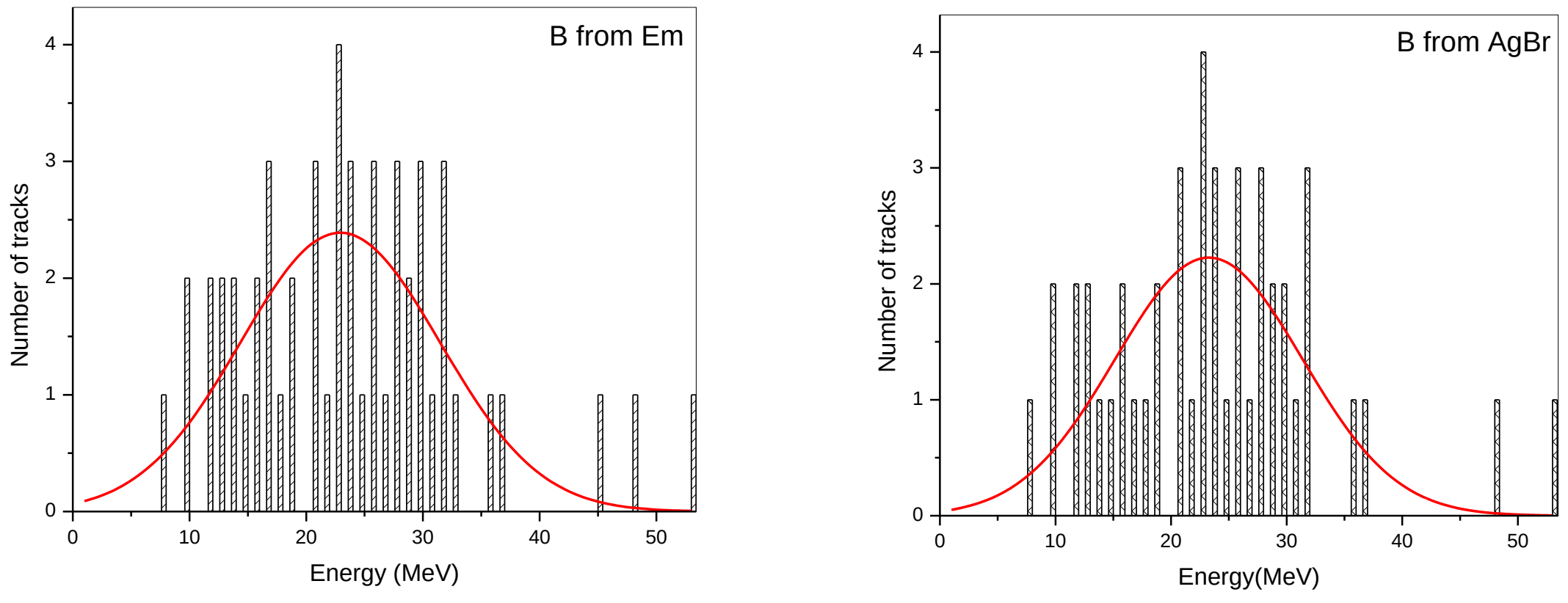

(e)

Fig. 4: Energy spectra of the emitted fragments from emulsion and Ag, Br targets. Histograms are experimental data while curves are their best fits.

In figure 4, one may notice that the main energy spectra of the emitted fragments are almost well fitted to Gaussian distributions with one dominant peak. The dependence of the peak position on the type of the fragment, as determined from the fitting parameters, is summarized in table 4. The errors refer to standard deviation according to figure (4).

Table 4: peak position and fragment type.

| Fragment | Peak position in (MeV) | | Peak position in (MeV) for the fragment/amu |
|---|---|---|---|
| | Em | Ag, Br | |
| P | 1.7$\pm$0.3 | 1.7$\pm$0.3 | 1.7$\pm$0.3 |
| He | 6.6$\pm$1.6 | 6.7$\pm$1.5 | 1.7$\pm$ 0.4 |
| $^{7}$Li | 12.8$\pm$3.1 | 12.7$\pm$3.0 | 1.8$\pm$ 0.4 |
| $^{9}$Be | 18.1$\pm$5.2 | 17.6$\pm$5.0 | 2.0$\pm$ 0.5 |
| $^{11}$B | 23.0$\pm$8.5 | 23.3$\pm$8.5 | 2.1$\pm$ 0.8 |

From figure 4 (a through e) and table 4, one may notice that the peak position of the fragment energy distribution shifts toward higher values as the fragment gets heavier. One may also observe that the peak position, for each fragment type, has almost the same value regardless of their source of emission, whether from all emulsion targets or from its heavy target group (Ag, Br). The last column in table 4 deals with some kind of scaling of the measured quantities. The scaling properties of some physical systems depend on the dynamical evolution of a large number of nonlinearly coupled subsystems and it was used here to explore the possibility that scaling phenomenon occurs in a thermodynamic nuclear system. It is clear that the peak positions of the energy distribution of the emitted fragment per atomic mass unit have the same value within the standard deviation, showing a constant scaling parameter over the range of fragment masses (1- 11 amu) of our data. This result could reasonably be accepted if one assumes that this energy expresses the boiling off temperature of each fragment type [33], regardless of the emitting target nucleus and the mass of the fragment. Some other peaks may be observed in the graphs that could be attributed to emission of the fragments in a pre-equilibrium stage and thus, their energies (temperatures) are some way, different from the corresponding values after the system has reached the equilibrium state. This result might agree with our analysis made to fig.3.

## Conclusion

The experimental data together with their related fittings and statistics of the electromagnetic interaction of relativistic muons with emulsion nuclei have shown a number of interesting results. The emission probability of alpha particle fragments has shown remarkable higher values than that of other fragments in the interaction of the muon with the heavy target nuclei as well as with all the emulsion constituent nuclei of. This confirms the fact that the alpha particle is a tightly bound structure and that nuclei are likely be described as clusters of alphas. The angular distributions of the emitted fragments have shown isotropic character, so that one would believe that all of these fragments could have been originated from similar processes in the excited nuclei. This trend is also seen in the emitted fragments of the residual excited nuclei in high energy hadrons and heavy ion strong interactions. The energy spectra of all types of the emitted fragments have shown different peak positions. These peaks are related to the "boiling off" energy (temperature) that characterizes each emitted fragment. This fact, together with the others, that have come out of this work will be exposed to extended analysis and studies and will be shown up in a series of forthcoming work in the near future.

## Acknowledgement

This work was supported by the Academy of Scientific Research and Technology (ASRT, Egypt) and the Joint Institute for Nuclear Research (JINR, Russia).

**REFERENCES**

[1] C. L. Timoshenko and V. M. Emelyanov, Fiz. Elem. Chastits At. Yadra 37, 1150 (2006) [Phys. Part. Nucl. 37, 606 (2006)].

[2] F. Krauss, M. Greiner, and G. Soff, Prog. Part. Nucl. Phys. 39, 503 (1997).

[3] G. Baur, K. Hencken, and D. Trautmann, J. Phys. G 24, 1657 (1998).

[4] G. Baur, K. Hencken, and D. Trautmann, S. Sadovsky, and Y. Khalrov, Phys. Rep. 364, 359 (2002).

[5] A. J. Baltz, G. Baur, D. d,Enterria, et. al., Phys. Rep. 458, 1 (2008).

[6] V. G. Nedorezov and Yu. N. Ranyuk, Fiz. Elem. Chastits At. Yadra 15, 379 (1984) [Sov. J. Part. Nucl. 15, 172 (1984)].

[7] J. Eisenberg and W. Greiner, *Nuclear Models* (North Holland, Amsterdam, 1988; Atomizdat, Moscow, (1875).

[8] B. S. Ishkhanov and V. N. Orlin, Fiz. Elem. Chastits At. Yadra 38, 460 (2007) [Phys. Part. Nucl. 38, 232 (2007)].

[9] P. S. Isaev, Quantum Electrodynamics at High Energies (Energoatomizdat, Moscow, 1984; Amer. Inst. Physics, New York, 1989).

[10] R. Palit, P. Adrich, T. Aumann, et al., Phys. Rev. C 68, 034318 (2003).

[11] C. A. Bertulani and V. Yu. Ponomarev, Phys. Rep. 321, 139 (1999).

[12] T. Aumann, P. F. Bortignon, and H. Emling, Ann. Rev. Nucl. Part. Sci. 48, 351 (1998).

[13] K. Boretzky, A. Grunschloss, S. Ilievski et al., Phys. Rev. C 68, 024317 (2003).

[14] A. Heinz, K. H. Schmidt, A. R. Junghans, et al. Nucl. Phys. A 713, 3 (2003).

[15] J. D. Jackson, Classical Electrodynamics, 2$^{nd}$ ed, Wiley, New York, 1975.

[16] E. Fermi, Nuovo Cim. 2, 143 (1925).

[17] E. Fermi, "On the Theory of Collisions between Atoms and Electrically Charged Particles", arXiv:hep-th/0205086.

[18] B. Hoffmann and G. Baur, Phys. Rev. C 30,247(1984).

[19] M. K. Singh et al., Indian J. Phys. 88, 323 (2014).

[20] Jun-Sheng Li et. al., Nucl. Instru. Meth. Phys. Res. B 307, 503 (2013).

[21] G. K. Mallot, The COMPASS collaboration- CERN, Nuclear Instruments and Methods in Physics Research Section A: Accelerators, Spectrometers, Detectors and Associated Equipment, 518, 121(2004)

[22] Dipak Ghosh, Argha Deb, Swarnapratim Bhattacharyya, Jayita Ghosh, Nuclear Physics A 720 (2003) 419

[23] D. E. Greiner et al., Phys. Rev. Lett. 35, 152 (1975).

[24] G. M. Chernov et al., Nucl. Phys. A 412, 534 (1985)

[25] V. A. Abdurazakova et al., Sov. J. Nucl. Phys. 47, 827 (1988).

[26] H. H. Hecknnan et al., Phys. Rev. C17, 1735 (1978).

[27] K. B. Bhalla, M. Chaudhry and S. L. Okanathan., Nucl. Phys. A367, 446 (1981).

[28] G. Singh, P. L., Jain, M. S. El-Nagdy, Europhys. Lett. A7, 113 (1992)

[29] M. EL-Nadi et al., Int. J. Mod. Phys. E.6, 191 (1997).

[30] Zhang Dong-Hai, Zhao Hui-Hua, Liu Fang, He Chun-Le, Jia Hui-Ming, Li Xue-Qin, Li Zhen-Yu and Li Jun-Sheng, Chin. Phys. Soc. 15, 9 (2006).

[31] LUKE C.L.YUAN and CHIEN-SHIUNG WU, Methods of Experimental Physics, Volume 5-Part A, p. 236, ACADEMIC PRESS INC. (LONDON), 1961.

[32] M. T. Hussein, N. M. Hassan, N. M. Sadek and J. Elsweedy. Fizika B (Zagreb) 15 (2006) 2, 51

[33] J. Toke, L. Pienkowski, L. Sobotka, M. Houck, and W. U. Schroder, Phys. Rev. C 72, 031601 (2005).